\documentclass[10pt, conference]{IEEEtran} 

\usepackage{graphicx}
\usepackage{multirow}
\usepackage{tabu}
\usepackage{array}
\usepackage{url}
\usepackage{amsmath}
\usepackage{amssymb}
\usepackage{amsthm}
\usepackage{amsfonts}
\usepackage{bm}
\usepackage{graphicx}
\usepackage{cite}
\usepackage{caption}
\usepackage{color}
\usepackage{subcaption}
\usepackage{amssymb}
\usepackage{tabulary}
\usepackage{booktabs}
\newtheorem{remark}{Remark}

\usepackage[justification=centering]{caption}
\IEEEoverridecommandlockouts                          

\title{
	 Performance Analysis of a System with Bursty Traffic and Adjustable Transmission Times
}

\author{
	\IEEEauthorblockN{Nikolaos Pappas
	\IEEEauthorblockA{Department of Science and Technology, Link{\"o}ping University, Sweden}
	E-mail: nikolaos.pappas@liu.se\\
	}
}

\begin{document}
	
\maketitle

\begin{abstract}
In this work, we consider the case where a source with bursty traffic can adjust the transmission duration in order to increase the reliability. The source is equipped with a queue in order to store the arriving packets. We model the system with a discrete time Markov Chain, and we characterize the performance in terms of service probability and average delay per packet. The accuracy of the theoretical results is validated through simulations. This work serves as an initial step in order to provide a framework for systems with arbitrary arrivals and variable transmission durations and it can be utilized for the derivation of the delay distribution and the delay violation probability.
\end{abstract}

	
\IEEEpeerreviewmaketitle
	
\section{Introduction}

One of the main goals of the next generation mobile communications is to provide seamless communication for a massive amount of devices building the Internet-of-Things (IoT) and at the same time to support the constantly increasing traffic demands originated from personal communications. The major difference between 5G and the previous generations is the native support of ultra-high reliability and low latency. These are required by several applications and services such as autonomous vehicles, factory automation, tele-presence, smart grids etc. The wireless traffic generated by these cases, often referred to as machine-type communication, is different from the traffic that can be supported efficiently by the current wireless communication systems, due to the stringent requirements in terms of latency and reliability \cite{durisi2016PIEEE}. The principles for supporting ultra-reliable low latency communication (URLLC) from the perspective of the traditional assumptions are discussed in \cite{PopovskiNETWORK2018}. Furthermore, that article elaborates on possible applications in various elements of system design, such as use of various diversity sources, design of packets, and access protocols. The work in \cite{NielsenTCOM2018} proposed interface diversity and integration of multiple communication interfaces in order to offer URLLC without intervention in the physical layer design.

Considering flexible transmission time interval (TTI) can be one option to provide low latency to services with strict latency requirements. In order to support services with heterogeneous requirements, the works in \cite{FlexiblePedersen} and \cite{FlexiblePedersen1} propose a flexible frame structure. In \cite{QiGCW2016}, scalable TTI lengths are introduced in order to consider the requirements of each individual service and provide a trade-off between heterogeneous performance metrics. The works in \cite{FountoulakisWiOpt2017} and \cite{You2018CL} develop scheduling approaches that fulfill the deadlines and requirements of different types of services by scaling the length of the used TTI. In \cite{DestounisWiOpt2018}, the authors propose a scheduling policy to activate users and uncertain short-packet transmissions with the goal to establish reliable latency performance.

The work in \cite{ChenWoWMoM2016, ChenTWC2018} considers the maximization of throughput under delay constraints in large scale wireless networks. In \cite{SPAWC2018}, the performance of deadline-constrained bursty traffic with retransmissions is studied under constant transmission time. In \cite{KountourisWiOpt2018}, the delay performance of large wireless networks in the presence of statistical QoS constraints is studied. The distribution of the conditional delay violation probability and effective capacity in Poisson bipolar networks has been characterized. A survey on the emerging technologies to achieve low latency communications can be found in \cite{ParvezCOMST}.

In this work, we consider a source with bursty traffic. The source can adjust the transmission duration based on a probabilistic model in order to increase the reliability. The source is equipped with a queue in order to store the arriving packets and the transmission is through an erasure wireless channel. Clearly, this work can be connected with the area of low latency communications and the transmission of short packets, since we can utilize the results from finite blocklength analysis regarding the error probability.
We aim to develop a framework by utilizing discrete time Markov Chains, and we characterize the performance in terms of service probability, stability conditions, and average delay per packet. 

The remainder of the paper is organized as follows. In Section \ref{sec:model}, we give the system model considered in this work. In Section \ref{sec:DTMC}, we provide the modeling based on a discrete time Markov Chain, in Section \ref{sec:approx}, we propose an approximation based on a Geo/Geo/1 queueing model which is simpler to analyze. In Section \ref{sec:res} we provide the simulation and numerical results and in Section \ref{sec:conclusions} we conclude our work and discuss future directions.

\section{System Model}\label{sec:model}
We are interested in studying a queueing system with slotted time. On each time slot we can have up to $N$ packet arrivals and up to $M$ packet departures. A general arrival model considers that $i$ packets arrive with probability $\alpha_i$ for $0 \leq i \leq N$ and $\sum_{i=0}^{N} \alpha_i=1$ during a timeslot. In addition, for the service process, we consider the case that the transmission time can be adjusted in order to occupy several time slots. We assume that we choose the duration for a transmission of a packet to occupy $j$ slots with probability $q_j$ for $1 \leq j \leq M$, with $\sum_{j=0}^{M} q_j=1$. By allowing the transmission of a packet to expand into several slots we can increase the probability of success; however, delay is increased as well. When $j$ timeslots are selected to transmit a packet then the success probability is $p_j$. In general we have $p_1 < p_2 < \dots <p_M$. In case of a transmission failure, then the duration of the next transmission will be decided independently. The ACKs are instantaneous and error free and all the packets have the same size. In addition, we assume a late arrival and early departure queueing model.

However, before proceeding with the general model described above, we will consider a simpler, yet complex enough, case to analyze. More specifically, in this work, we consider a source with fixed average arrival probability $\lambda$. The source can adjust probabilistically its transmission time between two options. With probability $q_1$, the source selects one slot, and with probability $q_2$ two slots, $q_1+q_2=1$. We could consider one more option, the case that the transmitter will not utilize a slot and it will remain silent, this could happen with probability $q_0$. This probability can be associated with the channel state if we assume that we can have this knowledge. However, this is outside of the scope for this work at this stage.

We assume that the transmission takes place over an erasure wireless channel. Thus, when the transmission duration is one timeslot the success probability is denoted by $p_1$; when the transmission lasts for two slots, then the success probability is $p_2$. We further assume that $0 \leq p_1 \leq p_2 \leq 1$. 
The model presented here can be connected with the transmission of short packets that is common in low latency communications. However, when we have transmission of
short packets, asymptotic information theoretic results do not apply. Thus, each transmission has a non-zero error probability, which can be approximated by \cite{PolyanskiyTIT2010}

\begin{equation} \label{eq:errorprob}
p_e (\gamma, b) \approx Q \left( \frac{n \log_2 (1+\gamma)-b+\frac{\log_2 (n)}{2}}{\sqrt{V(\gamma) n}}  \right),
\end{equation}

where $n$ is the number of channel uses, $b$ is the number of transmitted bits and it can be connected to the packet size, $\gamma$ is the signal-to-noise ratio. In addition, $Q(.)$ is the Gaussian Q-function, and $V(.)$ is the channel dispersion. More details on the error probability can be found in \cite{PolyanskiyTIT2010}. The success probabilities $p_1$ and $p_2$ in this model can be connected with the approximation in \eqref{eq:errorprob}. For this work we intent to keep the analysis quite general thus, we do not consider such connection.

The notation used in this work is summarized in Table \ref{table:notation}.

\begin{table}[!t]
	\small
	\caption{Notation}
	\centering
	\begin{tabular}{ | c | l | }
		\hline
		\bfseries Symbol & \bfseries Explanation\\
		\hline
		$\lambda$ & Arrival probability of a packet in a timeslot\\
		\hline$q_1$ & Probability that a packet transmission \\& will occupy one timeslot \\
		\hline$q_2$ &Probability that a packet transmission \\& will occupy two timeslots\\
		\hline$p_1$ &The success probability of a packet when \\& the transmission duration is one timeslot\\
		\hline$p_2$ & The success probability of a packet when \\& the transmission duration is two timeslots\\ 
		\hline
	\end{tabular} \label{table:notation}	
\end{table}

\section{Analysis} \label{sec:DTMC}

\begin{figure*}[ht!]
\centering
\includegraphics[scale=0.4]{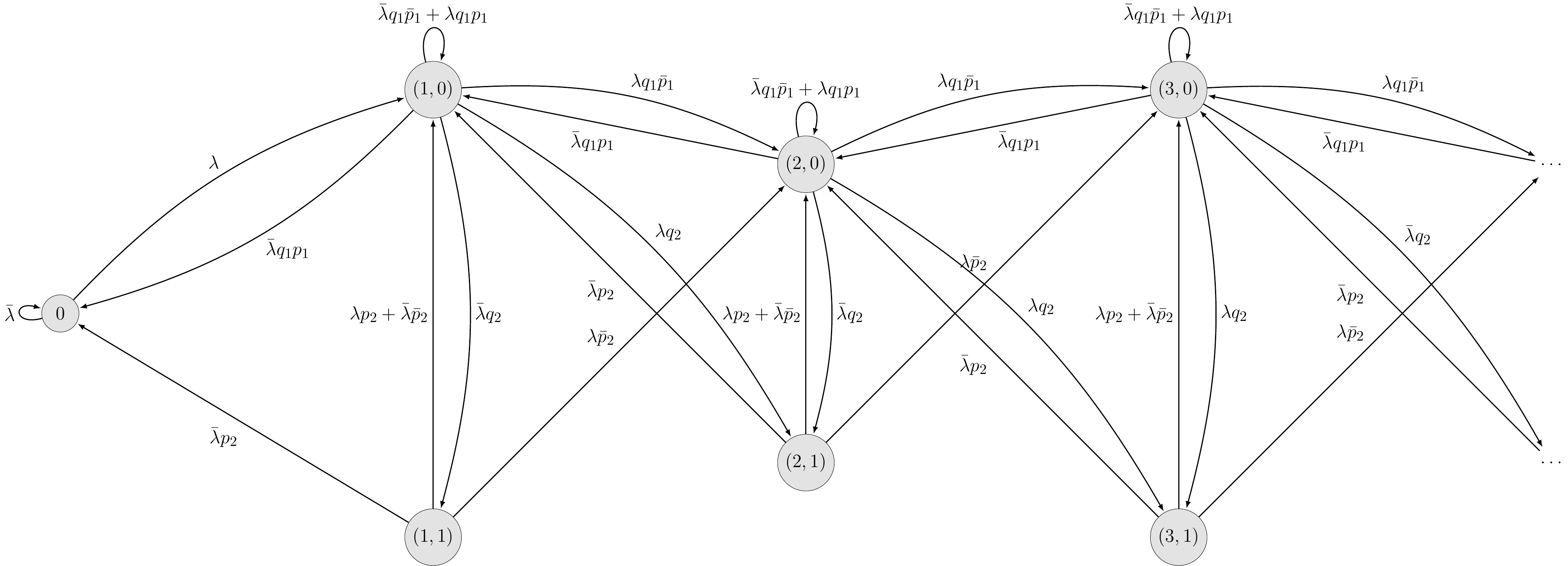}
\caption{The DTMC for the considered system. Note that $\bar{x}=1-x$.}
\label{fig:MC1}
\end{figure*}

In order to characterize the performance of the considered system, we model the queue evolution and the operation of the system as a Discrete Time Markov Chain (DTMC) with infinite number of states depicted in Fig. \ref{fig:MC1}. In the figures for the DTMC in this section and in the next one, we use the bar symbol to denote the complementary probability, $\bar{x} = 1-x$, for presentation reasons.
The state denoted by $0$ models an empty system, then we have two cases for the states, the states $(i,0)$ and $(i,1)$ for $i \geq 1$. The state $(i,0)$ denotes that there are $i$ packets in the queue and there is no packet in service from the previous slot, the state $(i,1)$ denotes that there are $i$ packets in the queue and there is a packet in service from the previous slot. 

Since we have the arrival of up to one packet with probability $\lambda$ the feasible transitions from state $0$ are to $(1,0)$ and to $0$.

\begin{figure}[ht!]
\centering
\includegraphics[scale=0.5]{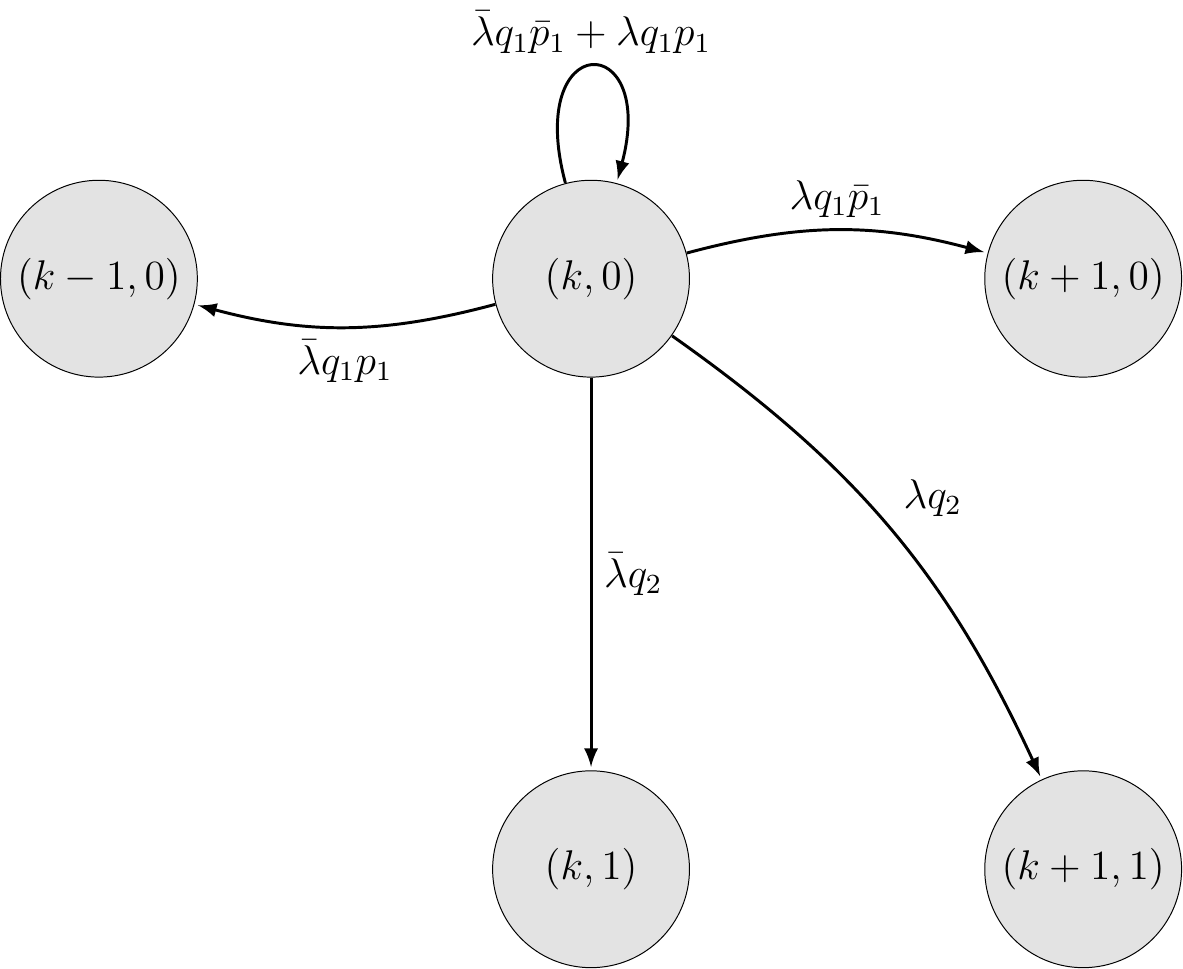}
\caption{The transitions and their probabilities from state $(k,0)$ for $k \geq 2$.}
\label{fig:MC11}
\end{figure}

The transitions from a state $(k,0)$ for $k\geq 2$ are depicted in Fig. \ref{fig:MC11}.
Note that the Markov Chain can remain in state $(k,0)$ either by not receiving a newly arrived packet and selecting one timeslot for transmission which fails, or a new packet arrives and the source selects one timeslot transmission which is successful. With similar reasoning one can obtain the other transition probabilities from $(k,0)$.

\begin{figure}[ht!]
\centering
\includegraphics[scale=0.5]{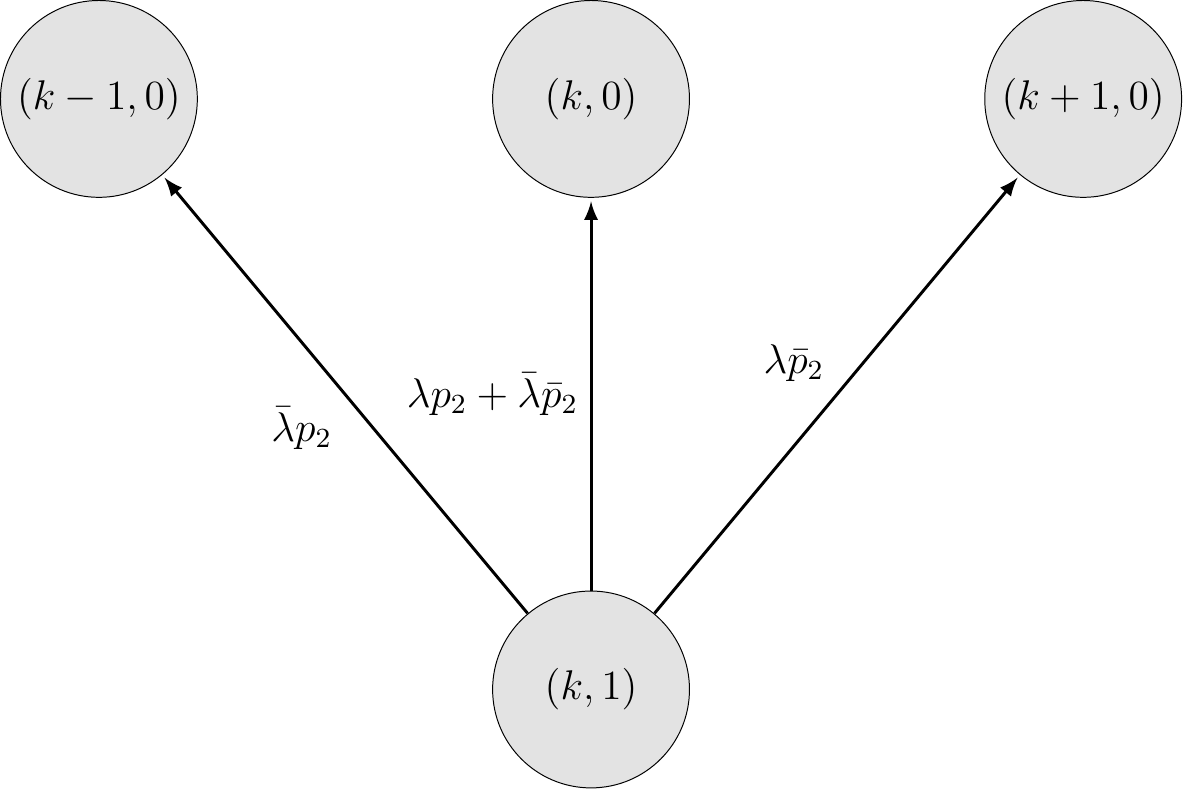}
\caption{The transitions and their probabilities from state $(k,1)$ for $k \geq 2$.}
\label{fig:MC12}
\end{figure}

The transitions from a state $(k,1)$ for $k\geq 2$ are depicted in Fig. \ref{fig:MC12}.
Note that this state denotes the case where there are $k$ packets stored in the queue and there is one under transmission because of a two-slot transmission duration. That packet will be successfully transmitted with probability $p_2$ at the end of the second slot.

For DTMCs with infinite states we can compute the stationary distribution vector $\pi$ by solving the system of equations $P\pi=\pi$ and $\sum_{i=0}^{\infty} \pi_i =1$, where $\pi_i$ is the $i$-th element of vector $\pi$. The transition matrix, $P$, is given by \eqref{eq:transmatrix}.

\begin{figure*}[ht]
\begin{align} \label{eq:transmatrix}
P = 
\begin{bmatrix} 
    \bar{\lambda} & \lambda & 0 & 0 & 0 & 0 & 0 & 0 & 0 & \dots \\
    \bar{\lambda}q_1p_1 & \bar{\lambda}q_1 \bar{p}_1+\lambda q_1 p_1 & \bar{\lambda}q_2 & \bar{\lambda}q_1 \bar{p}_1 & \lambda q_2 & 0 & 0 & 0 & 0 & \dots \\  
    \bar{\lambda}p_2 & \lambda p_2 +\bar{\lambda} \bar{p}_2 & 0 & \lambda \bar{p}_2 & 0 & 0 & 0 & 0 & 0 & \dots \\
    0 & \bar{\lambda}q_1p_1 & 0 & \bar{\lambda}q_1 \bar{p}_1+\lambda q_1 p_1 & \bar{\lambda} q_2 & \lambda q_1 \bar{p}_1 & \lambda q_2 & 0 & 0 & \dots \\   
    0 & \bar{\lambda} p_2 & 0 & \lambda p_2 + \bar{\lambda} \bar{p}_2 & 0 & \lambda \bar{p}_2 & 0 & 0 & 0 & \dots \\
    0 & 0 & 0 & \bar{\lambda}q_1p_1 & 0 & \bar{\lambda}q_1 \bar{p}_1+\lambda q_1 p_1 &  \bar{\lambda} q_2 & \lambda q_1 \bar{p}_1 & \lambda q_2 & \dots \\   
    \vdots & \vdots & \vdots & \vdots & \vdots & \vdots & \vdots & \vdots & \ddots & \dots \\   
    \end{bmatrix}
    \end{align}
\end{figure*}

From the transition matrix we observe that this Markov Chain has the structure of a Quasi-Birth-and-Death (QBD) DTMC and in order to find the stationary distribution we need to deploy semi-analytical methods such as the Matrix Analytical Methods. In general, for this type of DTMCs it is not easy to find closed form expressions. A detailed treatment on Matrix Analytical Methods can be found in \cite{alfa2016applied} and in \cite{harchol2013performance}.

In the next section we will consider a simpler model to approximate the behavior of this system by providing closed-form expressions.

\section{Approximation with a Geo/Geo/1 Queueing Model}
\label{sec:approx}
In this section, we will construct a system that approximates the performance of the previously described system. More specifically, we assume that the arrival probability is $\lambda$ and the average service probability in a timeslot is 
\begin{equation} \label{eq:mu1}
\mu=q_1p_1+\frac{q_2p_2}{2}.
\end{equation}
The second term in $\mu$, $\frac{q_2p_2}{2}$, is divided by two due to the fact that when we select the transmission over two slots, the average service probability we see over one slot is the half. We would like to clarify that this is a way to approximate the behavior of the previous system and it is much easier to analyze and provide closed-form expressions. Furthermore, in the next sub-section we will provide a better approximation for the service probability. In the next section we will evaluate the accuracy of this approximation.

\begin{figure}[ht!]
\centering
\includegraphics[scale=0.6]{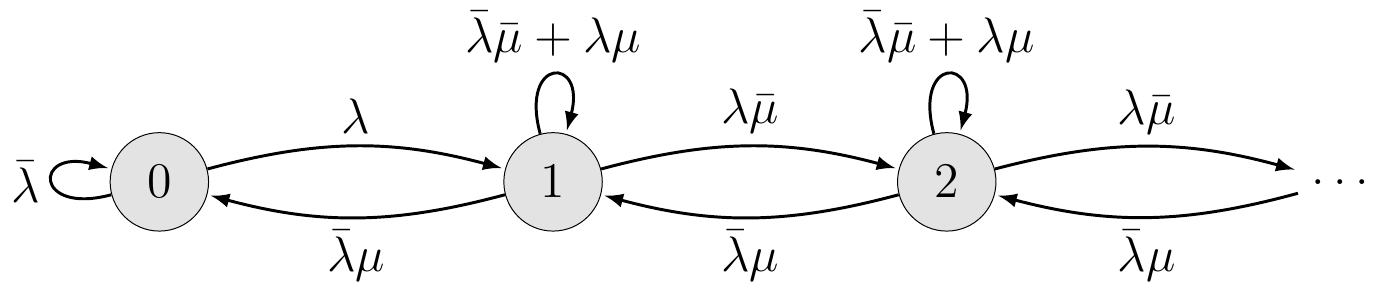}
\caption{The DTMC for the approximated model.}
\label{fig:MC2}
\end{figure}

The state diagram of the Discrete Time Markov Chain that describes the evolution of the new system is given in Fig. \ref{fig:MC2}. Recall that we assume an early departure and late arrival model. 

In order to compute the stationary distribution we utilize the balance equations. The stationary distribution of the DTMC is denoted by $\pi$, where $\pi(i) = \mathrm{Pr}\left(Q=i \right)$ is the probability that the queue has $i$ packets when it is in steady state.

From the balance equations we obtain the following
\begin{equation*}
\lambda \pi(0) = (1-\lambda) \mu \pi(1) \Leftrightarrow \pi(1) = \frac{\lambda}{(1-\lambda) \mu} \pi(0),
\end{equation*}
\begin{align*}
\left[\lambda (1-\mu) +(1-\lambda)\mu\right]\pi(1) = \lambda \pi(0)+(1-\lambda)\mu \pi(2) \\
 \Leftrightarrow \pi(2) = \frac{\lambda^2 (1-\mu)}{(1-\lambda)^2 \mu^2} \pi(0).
\end{align*}

Similarly, for $i > 1$ we have 
\begin{equation*}
\pi(i) = \frac{\lambda^i (1-\mu)^{i-1}}{(1-\lambda)^i \mu^i} \pi(0).
\end{equation*}

The previous steady state probabilities are given as a function of $\pi(0)$, however 
\begin{equation} \label{eq:sum}
\sum_{i=0}^{\infty} \pi(i)=1.
\end{equation}

Then we can obtain the probability that the queue is empty and is given by
\begin{equation}
\mathrm{Pr}\left(Q=0 \right) = 1-\frac{\lambda}{\mu}.
\end{equation}

\begin{remark}
The stability condition of the queue is $\frac{\lambda}{\mu} <1$.
Then we can obtain the values of $q_1$ that give a stable queue. 
If $p_1 > \frac{p_2}{2}$ then $q_1 > \frac{2\lambda-p_2}{2p_1-p_2}$, else if $p_1 < \frac{p_2}{2}$ then $q_1 < \frac{2\lambda-p_2}{2p_1-p_2}$. The stability of a queue implies finite queueing delay.
\end{remark}
\begin{remark}
Recall that when a queue is stable, then the throughput, which is also called stable throughput, is the arrival probability. If the queue is unstable, then the throughput is the service probability, and it is also called saturated throughput.
\end{remark}

The average queue size, $\bar{Q},$ can be computed by
\begin{equation}
\bar{Q}=\sum_{i=1}^{\infty} i\pi(i)=\frac{\lambda(1-\lambda)}{\mu - \lambda}.
\end{equation}

\subsection{Average Delay}

The delay per packet consists of the queueing delay and the transmission delay. From Little’s law, we obtain the queueing delay, $D_Q$, which is related to the average queue size per packet arrival and is
\begin{equation} \label{eq:DQinit}
D_Q=\frac{\bar{Q}}{\lambda}=\frac{1-\lambda}{\mu-\lambda}.
\end{equation}

The transmission delay, $D_T$, can be found by applying the regenerative method \cite{Walrand} as follows.

When a packet is transmitted from the source and the selected duration is one timeslot there is a probability that this packet reaches the destination during this slot, which is $q_1 p_1$. If a transmission duration of two slots is selected with probability $q_2$, then at the end of the second (next) slot there is a probability $p_2$ that the packet reaches the destination successfully. In both cases, if the transmission to the destination
is not successful, then the packet remains in the queue and it will be retransmitted in the next slot. In the next slot the decision of the transmission duration is independent of the past.

Then, we have that
\begin{equation} 
D_T=q_1 p_1 + q_1 (1-p_1) (1+D_T) + q_2(1+D_2),
\end{equation}
where, $D_2=p_2+(1-p_2)(1+D_T)$. Then we obtain that the transmission delay is

\begin{equation} \label{eq:DT}
D_T=\frac{q_1+2q_2}{1-q_1(1-p_1)-q_2(1-p_2)}.
\end{equation}

Consider the two extreme cases, the first is $q_1=1$ then $D_T=\frac{1}{p_1}$, and the second is $q_2=1$ then $D_T=\frac{2}{p_2}$.

The expression for the transmission delay in \eqref{eq:DT} can be also written as 

\begin{equation} \label{eq:DTq1}
D_T=\frac{2-q_1}{p_2+q_1(p_1-p_2)}.
\end{equation}

\begin{remark}
The transmission delay obtained in \eqref{eq:DT}, is the exact one since we didn't use the approximated service probability of the queue presented previously. Thus, here we can work in an inverse way to obtain a better approximation for $\mu$. Since, we know that the transmission delay can be connected with the service probability by $D_T=\frac{1}{\mu}$, then we have that 
\begin{equation}\label{eq:muDT}
\mu=\frac{p_2+q_1(p_1-p_2)}{2-q_1}.
\end{equation}
As we will see in Section \ref{sec:res}, this is a quite accurate approximation.
The important observation here is that we worked in the opposite way, since we first characterized the transmission delay and then we obtained the service probability. Following the same methodology we can derive the service probability in the more general case where the transmitter can choose up to $M$ slots to transmit. 
\end{remark}

Based on $\frac{\lambda}{\mu}<1$ we can obtain more accurate stability conditions than the one in Remark 1.

After replacing \eqref{eq:muDT} in \eqref{eq:DQinit} we obtain that the queueing delay is given by

\begin{equation} \label{eq:DQ}
D_Q=\frac{(1-\lambda)(2-q_1)}{p_2+q_1(p_1-p_2)-\lambda(1-q_1)}.
\end{equation}

We can consider the optimization problem of minimizing the transmission delay over $q_1$.
If we utilize the expression in \eqref{eq:DTq1}, we can obtain that the optimal values of $q_1$ can be summarized in Table \ref{tab:minDT}.

\begin{table}[ht]
\begin{center}
		\renewcommand{\arraystretch}{1.5}
		\caption{The optimal values of $q_1$ that minimize the transmission delay $D_T$.}
\begin{tabular}{|c|c|c|}
\hline
 & $D_T^*$ & $q_1^*$ \\\hline
$p_1= \frac{p_2}{2}$ & $\frac{2}{p_2}$ & 0.5  \\ \hline
$p_1> \frac{p_2}{2}$ & $\frac{1}{p_1}$ & 1  \\ \hline
$p_1< \frac{p_2}{2}$ & $\frac{2}{p_2}$ & 0 \\ \hline
\end{tabular}
\label{tab:minDT}
\end{center}
\end{table}


In addition it will be of interest to minimize the average total delay for a given $\lambda$ that satisfies the stability conditions.

\begin{eqnarray}
\underset{q_1} {\min} ~&& D_Q+D_T\\
\mbox{s.t.} 
&&   0 \leq q_1 \leq 1 . \nonumber\\
&&  \text{Stability Conditions}   \nonumber.
\end{eqnarray}

In the next section, we will evaluate numerically this optimization problem.

\section{Numerical and Simulation Results}
\label{sec:res}
In this section we evaluate the performance of the considered system. We construct a Matlab-based behavioral simulator and we also evaluate numerically the system based on the analysis above. The simulations were performed for $10^6$ slots. In the presented plots when we refer to approximation we mean the case where the service probability is approximated by \eqref{eq:mu1}. Approximation 2 refers to the case where the service probability is given by \eqref{eq:muDT}.

\subsection{Service Probability}

Here we evaluate the accuracy of the proposed approximations regarding the service probability $\mu$ for several values of $p_1$, $p_2$, and $q_1$. 

\begin{figure}[ht!]
\centering
\includegraphics[scale=0.5]{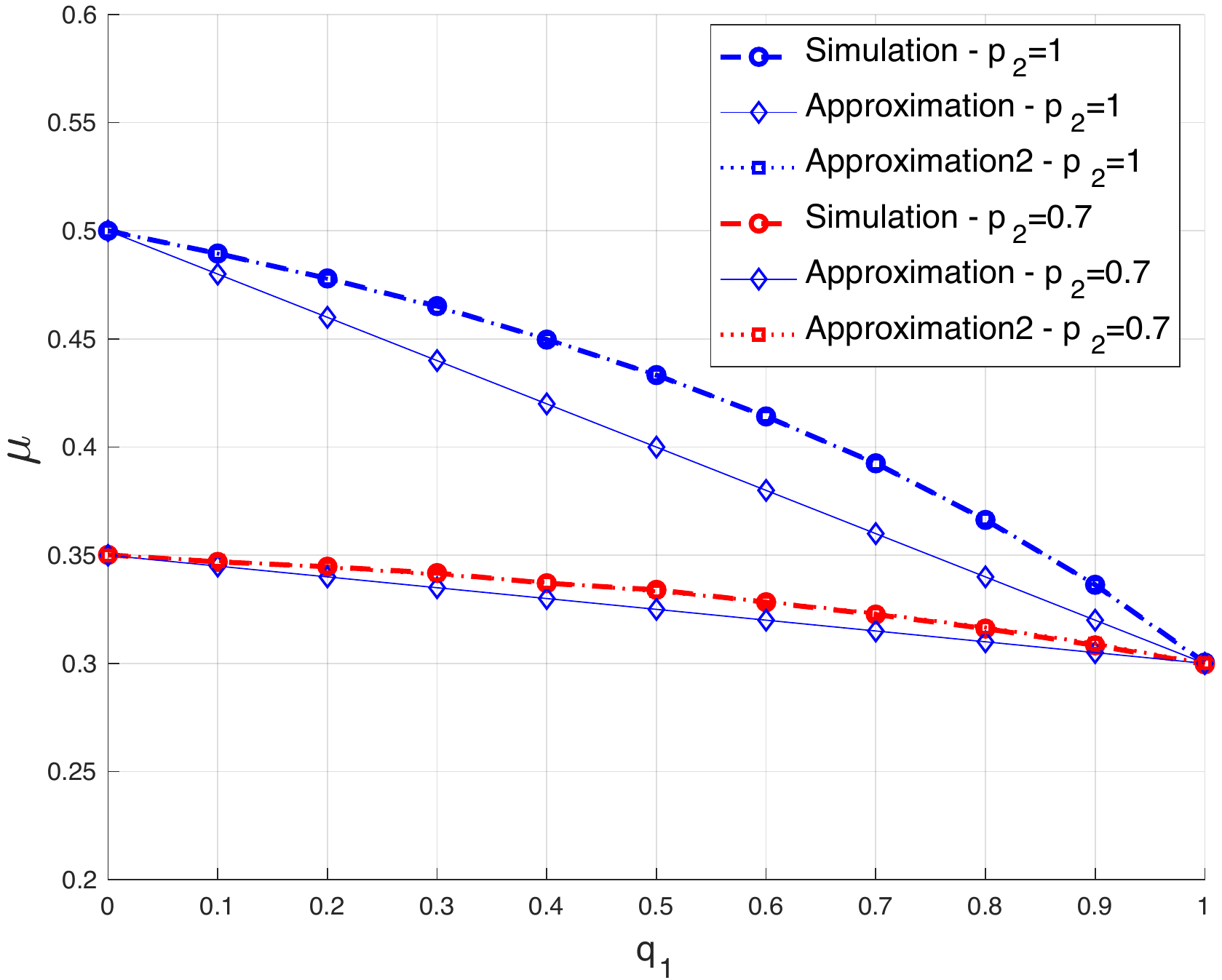}
\caption{$\mu$ versus $q_1$ for $p_1=0.3$.}
\label{fig:mu1}
\end{figure}

The biggest deviation for approximation based on \eqref{eq:mu1} is less than $8\%$.
We observe that for the presented cases, the approximated service probability from \eqref{eq:muDT} is the same with the simulated one. 

\begin{figure}[ht!]
\centering
\includegraphics[scale=0.5]{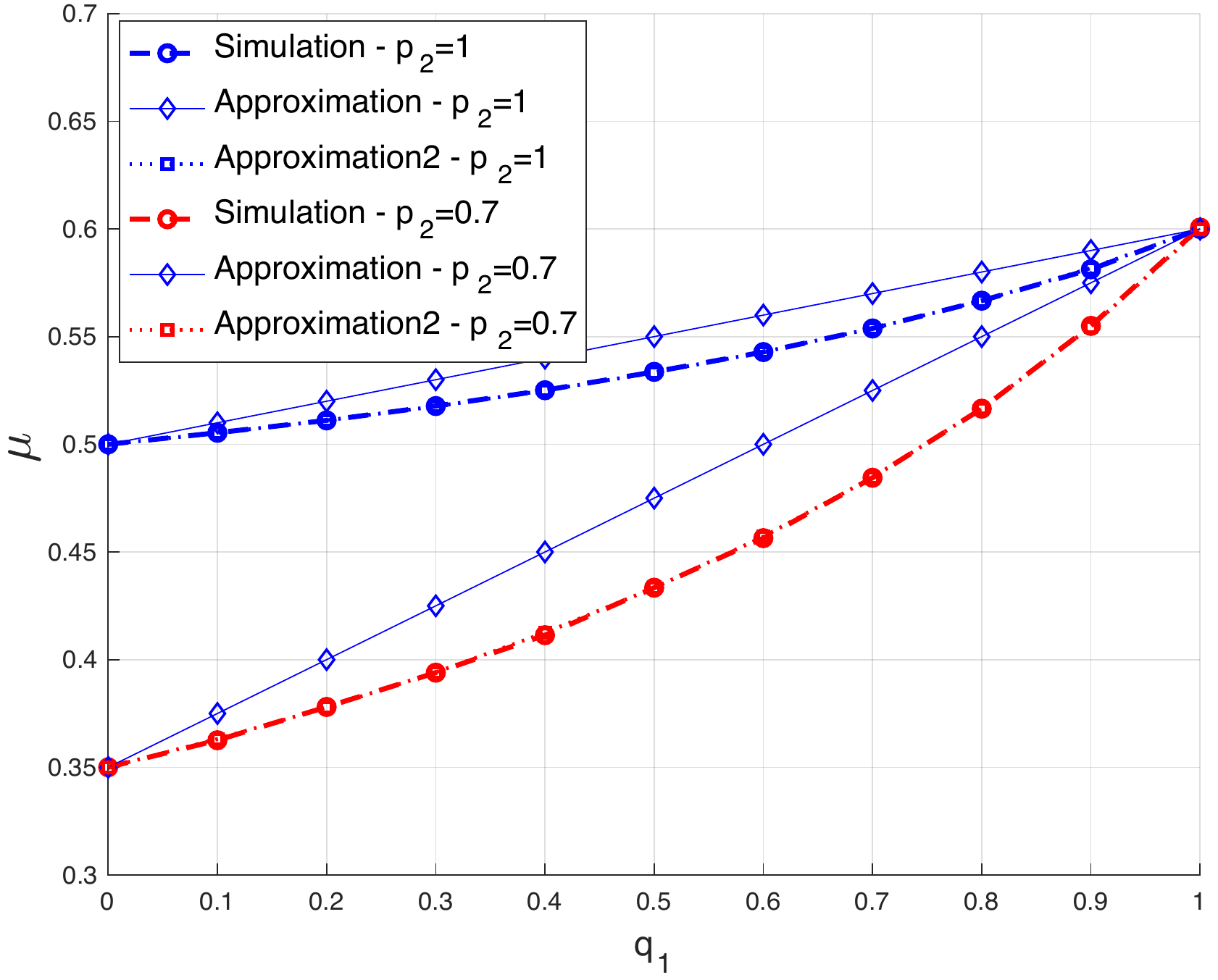}
\caption{$\mu$ versus $q_1$ for $p_1=0.6$.}
\label{fig:mu2}
\end{figure} 

The second approximation characterizes accurately the performance of the system regarding the service probability and is much simpler to analyze compared with the DTMC presented in Section \ref{sec:DTMC}.

\subsection{Average Delay}
\begin{figure}[ht!]
\centering
\includegraphics[scale=0.5]{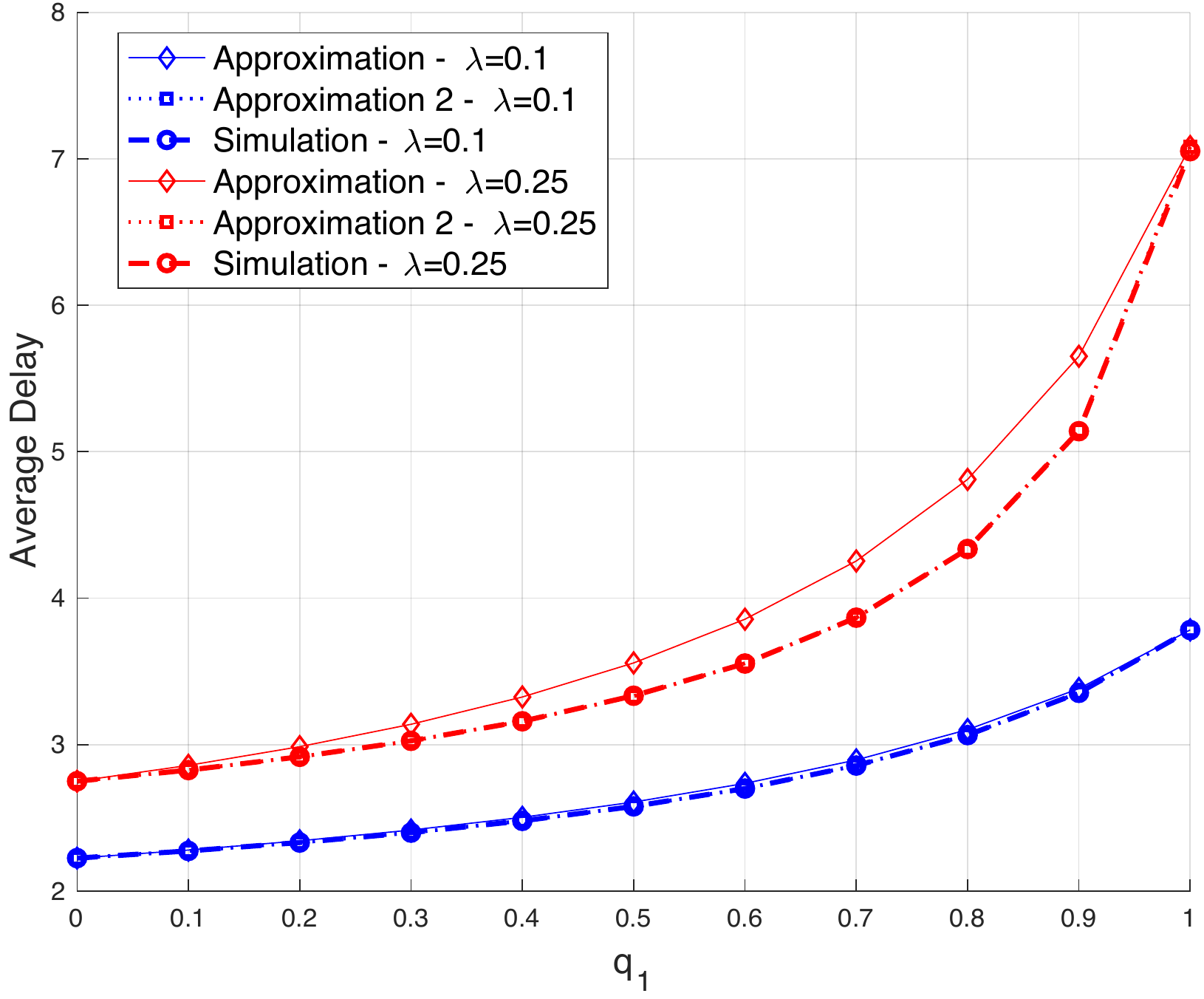}
\caption{Average Delay versus $q_1$ for $p_1=0.3$.}
\label{fig:delay1}
\end{figure}

\begin{figure}[ht!]
\centering
\includegraphics[scale=0.5]{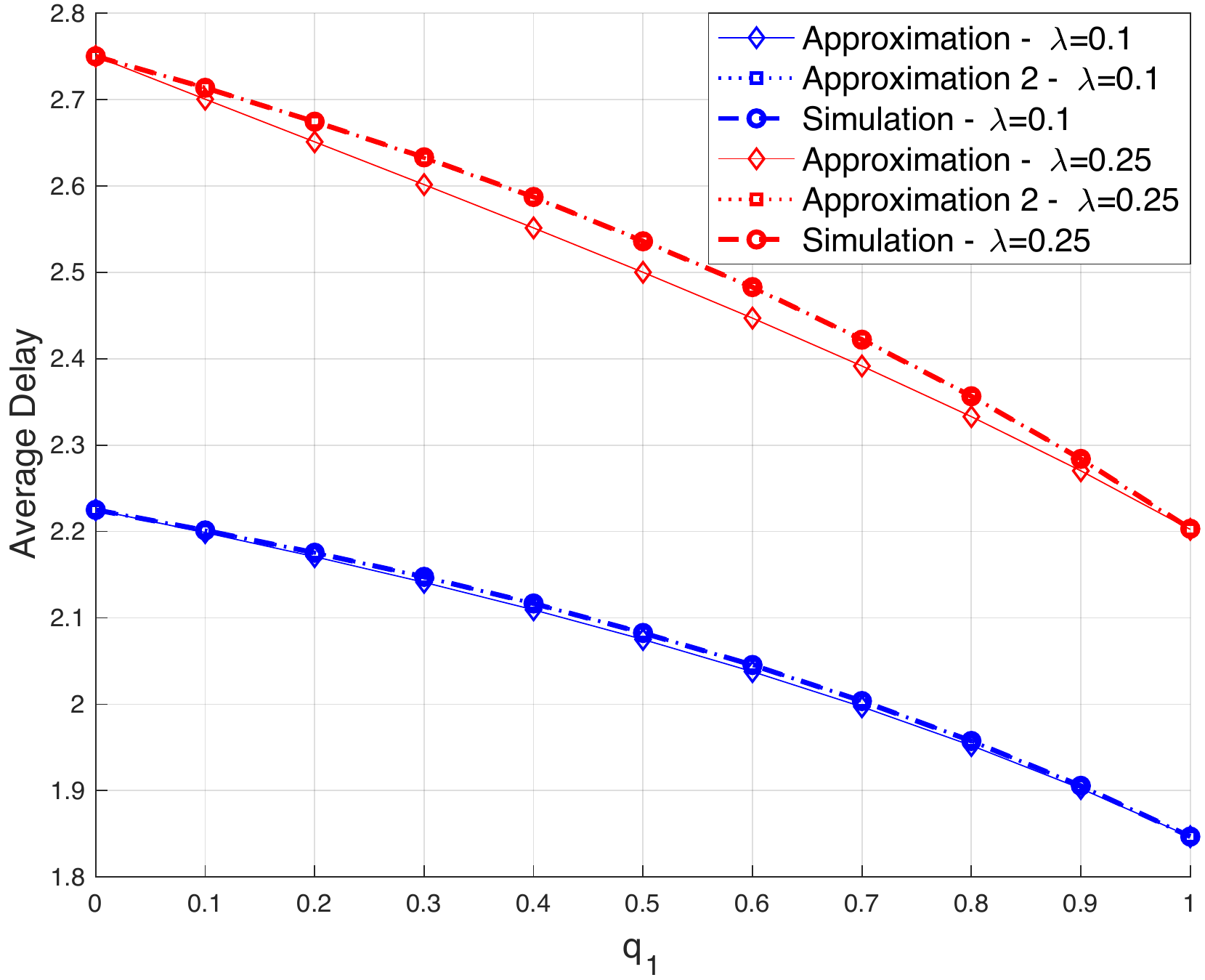}
\caption{Average Delay versus $q_1$ for $p_1=0.6$.}
\label{fig:delay2}
\end{figure}

Here we present the performance in terms of the average delay seen by a packet measured in timeslots. We consider two cases for the success probability for the one slot transmission, when the success probability is low, $p_1=0.3$, and when is higher, $p_1=0.6$, depicted in Fig. \ref{fig:delay1} and Fig. \ref{fig:delay2} respectively. For both cases we assume that $p_2=1$. Furthermore, we evaluate the accuracy of the two approximation models by comparing the delay from the approximation models with the delay obtained from simulation.

Figure \ref{fig:delay1} presents the average delay versus $q_1$ for two cases of arrival probability $\lambda=0.1$ (low-medium traffic regime) and $\lambda=0.25$ (medium-high traffic regime for this setup) when $p_1=0.3$. The average delay is increasing with the increase of $q_1$, this is expected, since it is preferable to have a two-slot transmission which it can compensate the frequent retransmissions. When $\lambda=0.25$ and $p_1=0.3$ then as $q_1$ increases then the queue tends to be unstable. In this case, the required retransmissions due to the high probability of failure is crucial for delay. The value of $q_1$ that minimizes the total delay Fig. \ref{fig:delay1} is $q_1^*=0$.

Figure \ref{fig:delay2} presents the average delay when $p_1=0.6$. In this case. it is better for the delay to have one slot transmission, since the success probability is high enough to compensate with the loss of one timeslot in case of two-slot transmission.
The value of $q_1$ that minimizes the total delay Fig. \ref{fig:delay2} is $q_1^*=1$.

The biggest deviation for the first approximation is less than $5\%$, so the proposed approximation model performs relatively well also regarding the delay. The observed discrepancy is caused by the queueing delay since the obtained transmission delay is the exact one.
\textit{The results from the second approximation coincide with the simulations which is expected as we discussed in the previous section in Remark 3}.

\section{Conclusions} \label{sec:conclusions}
In this work, we considered the case where a source, with bursty traffic stored in a queue, can adjust the transmission duration in order to increase the reliability. We modeled the system with a discrete time Markov Chain, and we characterized the performance in terms of service probability and average delay per packet. The accuracy of the theoretical results were validated through simulations. This is an initial study, the goal is to provide a framework for more general setups. The results in this work can be utilized for the derivation of the delay distribution; then other useful metrics, such as the delay violation probability can be obtained.

Future extensions include the general case with up to $N$ packet arrivals and up to $M$ timeslots duration for a packet transmission. Optimizing the selection probabilities for the duration of the transmission based on the traffic characteristic, the queue backlog and the state of the channel is important. Consideration of traffic with deadlines and power control in such a setup is a future step. Another crucial parameter is the consideration of hybrid automatic repeat request (HARQ) between the transmissions.

\section*{Acknowledgment}
This work was supported in part by the Center for Industrial Information Technology (CENIIT), ELLIIT, and the EU project DECADE under Grant H2020-MSCA-2014-RISE: 645705, the European Union’s Horizon 2020 research and innovation programme.

\bibliographystyle{IEEEtran}
\bibliography{bibliography}

\end{document}